\documentclass[a4paper,11pt]{article}
\usepackage{pos}
\usepackage{subfig}
\usepackage{wrapfig}
\usepackage{siunitx}
\usepackage{lipsum} 
\usepackage[capitalise]{cleveref}
\usepackage{graphicx}
\usepackage{lineno}

\title{Density of GeV Muons Measured with IceTop}
 \ShortTitle{Density of GeV Muons Measured with IceTop}

\author{The IceCube Collaboration \\{\normalsize \normalfont(a complete list of authors can be found at the end of the proceedings)}}
\emailAdd{soldin@udel.edu}

\abstract{We present a measurement of the density of GeV muons in near-vertical air showers using three years of data recorded by the IceTop array at the South Pole. We derive the muon densities as functions of energy at reference distances of 600\,m and 800\,m for primary energies between 2.5\,PeV and 40\,PeV and between 9\,PeV and 120\,PeV, respectively, at an atmospheric depth of about $690\,\mathrm{g/cm}^2$. The measurements are consistent with the predicted muon densities obtained from Sibyll~2.1 assuming any physically reasonable cosmic ray flux model. However, comparison to the post-LHC models QGSJet-II.04 and EPOS-LHC shows that the post-LHC models yield a higher muon density than predicted by Sibyll 2.1 and are in tension with the experimental data for air shower energies between 2.5\,PeV and 120\,PeV.

\vspace{4mm}
{\bfseries Corresponding author:}
Dennis Soldin$^{1*}$\\
{$^{1}$ \itshape Bartol Research Institute, Dept. of Physics and Astronomy\\
University of Delaware, Newark, DE 19716, USA}\\
$^*$ Presenter
\FullConference{37$^{\rm{th}}$ International Cosmic Ray Conference (ICRC 2021)\\
		July 12th -- 23rd, 2021\\
		Online -- Berlin, Germany}

}

\begin{document}
\maketitle

%\linenumbers

\section{Introduction}
\label{sec:intro}

Cosmic rays interact in the Earth’s atmosphere and produce extensive air showers (EAS) which can be measured with large detector arrays at the ground. The properties of the initial cosmic ray, such as energy and mass, are inferred indirectly from the particles measured at the ground and their interpretation strongly relies on simulations of the EAS development and thus on theoretical models~\cite{Kampert:2012mx}. One of the main challenges in understanding EAS is the description of hadronic interactions over several decades in center-of-mass energy. The relevant interactions are in the forward fragmentation region which can not be studied with existing colliders and their cross-sections cannot be computed from perturbative quantum chromodynamics. Instead, they are calculated using phenomenological models tuned to a variety of data sets from collider and fixed-target experiments, and are extrapolated into the phase space relevant for EAS. Several hadronic interaction models are available where the most recent ones take high-energy data from the LHC into account and are thus commonly referred to as \emph{post-LHC} models, such as EPOS-LHC~\cite{Pierog:2013ria} and QGSJet-II.04~\cite{Ostapchenko:2010vb}, in contrast to older \emph{pre-LHC} models, like Sibyll~2.1~\cite{Ahn:2009wx}. Air shower experiments can test and help to improve hadronic models with measurements of the muon content in EAS. 

This article reports a measurement of the density of muons with energies of around \SI{1}{GeV} at large lateral distances in EAS with energies between \SI{2.5}{PeV} and \SI{120}{PeV} with IceTop. The results will be compared to predictions from simulations based on recent hadronic interaction models and discussed in the context of different cosmic ray mass composition assumptions.% based on recent experimental measurements.

\section{IceTop}

\begin{wrapfigure}{r}{0.54\textwidth}
\vspace{-1.5em}
\mbox{\hspace{-.4em}\includegraphics[width=0.6\textwidth]{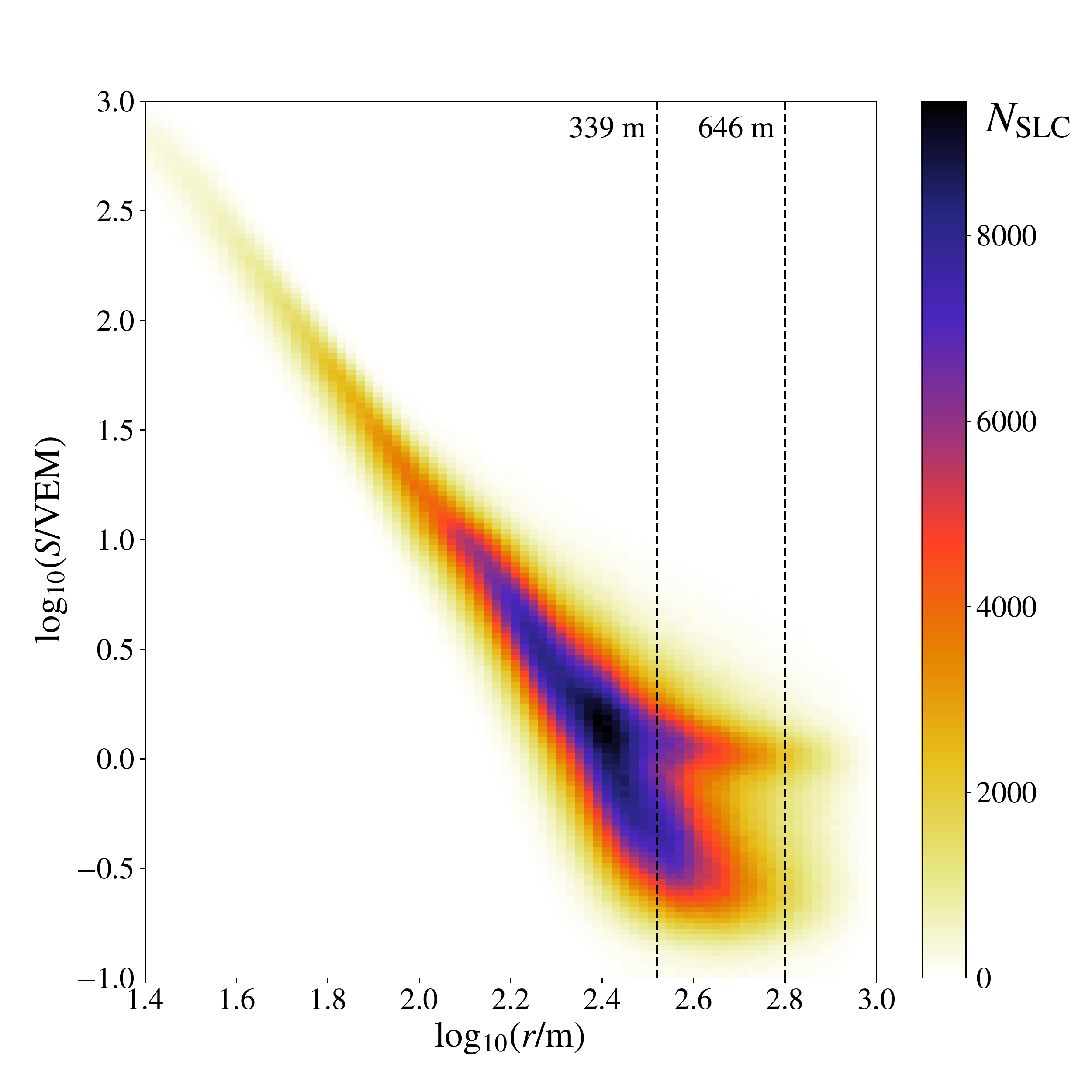}}
\vspace{-1.5em}
  
\caption{Distribution of SLC signals for near-vertical events ($\theta < \SI{18}{\degree}$) with reconstructed energies between \SI{10}{PeV} and \SI{12.5}{PeV} as a function of lateral distance and charge. A characteristic structure produced by muons in IceTop tanks at large distances, the so-called \emph{Muon Thumb}, is visible at signals around \SI{1}{VEM}.}
\vspace{-.5em}
\label{fig:thumb_plot}

\end{wrapfigure}

IceTop~\cite{IceCube:2012nn} is the surface air shower detector of the IceCube Neutrino Observatory~\cite{Aartsen:2016nxy} which is located at an altitude of about \SI{2.8}{km} above sea level (average atmospheric depth of $692 \,\mathrm{g/cm}^2$) at the geographic South Pole. It consists of $81$ stations deployed in a triangular grid with a typical separation of \SI{125}{m}. Each station consists of two tanks separated by about \SI{10}{m} which are filled with clear ice and contain two Digital Optical Modules (DOMs) which measure the Cherenkov light produced by EAS particles traversing the tanks. These stations have two readout modes: a Hard Local Coincidence (HLC) hit occurs when both tanks in a station have a discriminator trigger within a time window of \SI{125}{\micro s}. If there is a discriminator trigger in only one tank, it is called a Soft Local Coincidence (SLC) hit. While for HLC hits the full wave form of the DOM signals is recorded, for an SLC hit only the integrated signal charge and a timestamp are available. The tank signals are calibrated to be expressed in units of Vertical Equivalent Muon (VEM), which is the average charge produced by a vertically through-going muon. To determine the number of muons in an EAS, only the SLC information is used because at large distances from the shower axis, where muons are expected to dominate the EAS content, mostly only one tank of a station is hit. %constitutes the tank’s signal which is calibrated to be expressed in units of Vertical Equivalent Muon (VEM), which is the average charge produced by a vertically through-going muon. 
The distribution of SLC hits for near-vertical events with reconstructed energies between \SI{10}{PeV} and \SI{12.5}{PeV} as a function of lateral distance from the shower core and charge is shown in~\cref{fig:thumb_plot}. While the electromagnetic shower component produces a signal distribution which approximately follows a power law over the entire lateral distance range, a characteristic structure at signals around \SI{1}{VEM} produced by muons becomes visible at large distances, the so-called \emph{Muon Thumb}. This population consists mostly of tanks hit by one muon, and it is used to determine the muon content in EAS.%, as presented in~\cref{sec:analysis}.

\vspace{3em}

The HLC hits are used to determine the shower direction, the intersection point of the shower axis with IceTop (the \emph{shower core}), and the \emph{shower size}. This is done by fitting the measured signals with a Lateral Distribution Function (LDF) and their times with a phenomenological model of the shower front, as described in Ref.~\cite{IceCube:2012nn}. The LDF includes an attenuation factor which accounts for the snow coverage on top of each tank. The energy of the primary cosmic ray, $E_\mathrm{reco}$, is estimated based on the shower size, $S_{125}$, defined as the signal at a lateral distance of \SI{125}{m}, and the true primary energy, as obtained from simulations based on Sibyll 2.1. The resulting energy resolution is better than $0.1$ in $\log_{10}(E_\mathrm{reco})$ for all energies considered. Only events with reconstructed energy $E_\mathrm{reco} \geq 2.5\,\mathrm{PeV}$ are considered, an energy above which IceTop reaches a detection efficiency close to $100\%$ for all cosmic ray masses, from hydrogen up to iron~\cite{Aartsen:2013wda}.

%\vspace{5em}

\section{GeV Muon Analysis}
\label{sec:analysis}

This analysis uses data collected by IceTop between May 31, 2010 and May 2, 2013 with more than $18$ million events which pass the selection criteria, corresponding to around $947$ days of data acquisition. The event selection only considers events with triggers recorded in more than $5$ stations and a successful EAS reconstruction. In addition, the shower core must be within the geometrical area of IceTop, the tank with the largest signal must not be at the edge of the array, and there must be at least one station with signal greater than \SI{6}{VEM}. This analysis is further restricted to near-vertical events with zenith angles $\theta<18^\circ$ in order to select near-vertical muons.

\begin{figure}[t]
  \centering
  \vspace{-2.3em}
  
  \subfloat[]{%
    \includegraphics[width=0.46\textwidth,trim=0.0 2.0em 0.0 0.0, clip]{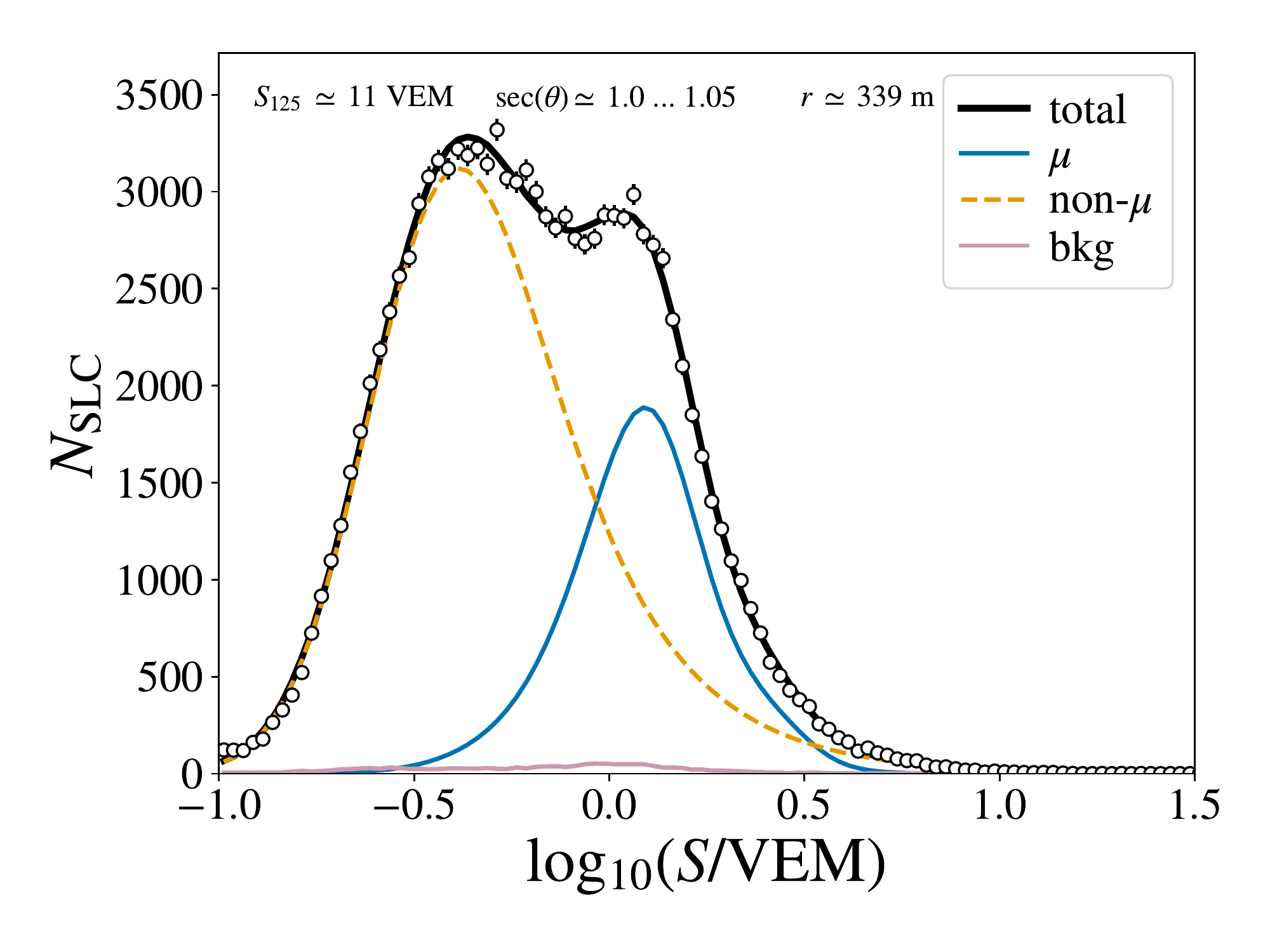}%
    \label{fig:slice_257m}%
  }\;\;
  \subfloat[]{%
    \includegraphics[width=0.46\textwidth,trim=0.0 2.0em 0.0 0.0, clip]{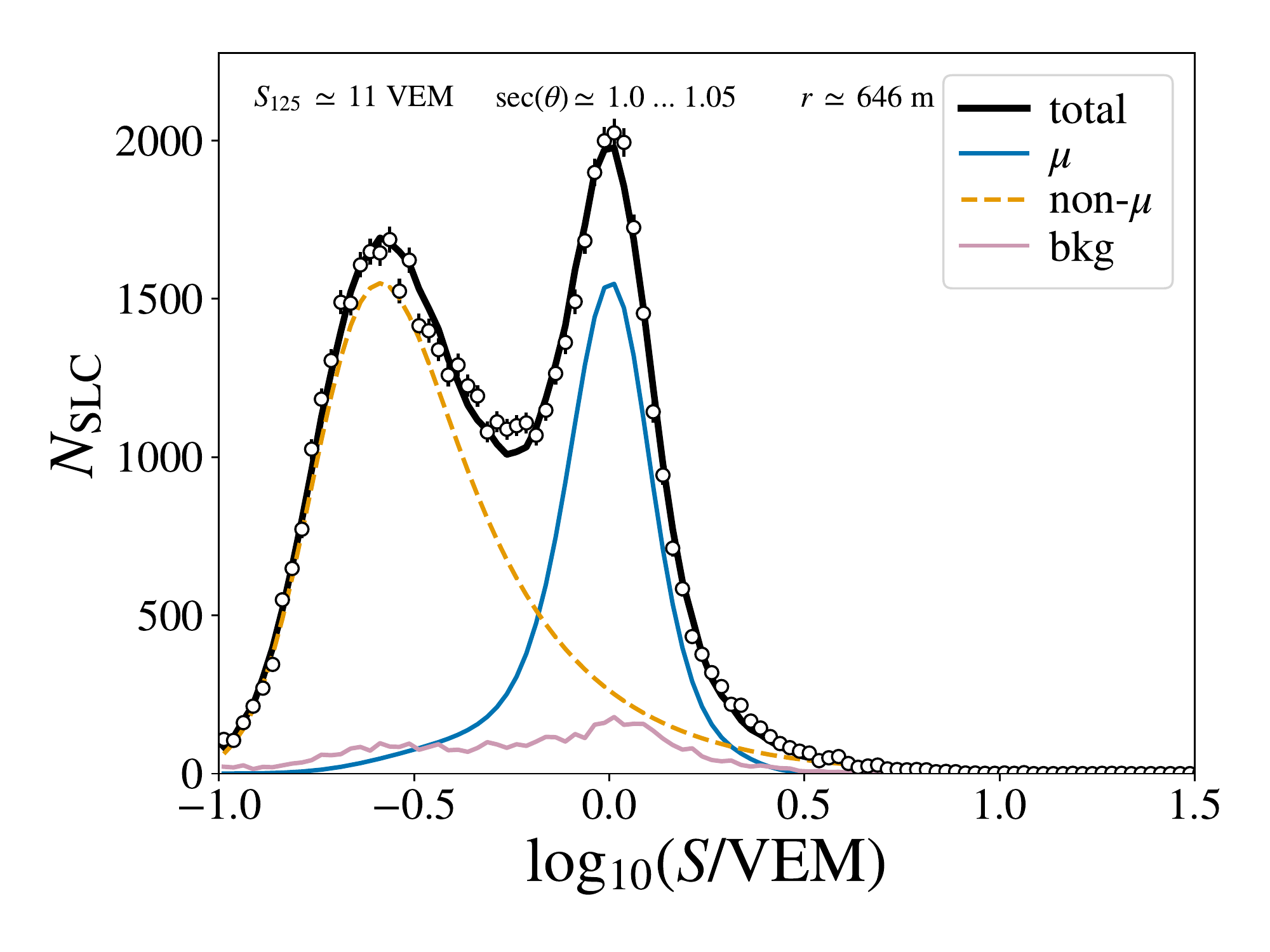}%
    \label{fig:slice_646m}%
  }
  \vspace{-0.6em}
  
  \caption{Signal distribution at lateral distances of \SI{339}{m} (a) and \SI{646}{m} (b), with fits to the signal model. These figures correspond to vertical slices in \cref{fig:thumb_plot}. The lines show the muon signal model (blue solid), the distribution of signals with no muons (dashed yellow) and the distribution of accidental signals (pink solid).}
  \label{fig:r_signal_slices}%
  \vspace{-1.1em}
  
\end{figure}

The muon content in EAS is determined based on the characteristic Muon Thumb in the signal distribution, shown in \cref{fig:thumb_plot}. This is done using a log-likelihood method to fit the signal distributions at fixed energy, zenith, and lateral distance (i.e. slices in $r$ of \cref{fig:thumb_plot}), using a multi-component model, which is illustrated in~\cref{fig:r_signal_slices} for lateral distances of about \SI{340}{m} and \SI{650}{m}. The figures show the muon signal distribution, an empirically determined distribution of signals with no muons, and the distribution of accidental signals. Thus the model includes individual signal models for the detector response to muons, the electromagnetic (EM) part of the EAS, and the contribution from accidental coincident background hits, which are described in detail in Refs.~\cite{Aartsen:2021Muons,Dembinski:2015xtn,Gonzalez:2019epd}. As shown in~\cref{fig:r_signal_slices}, the muon peak becomes dominant at large lateral distances where a large fraction of the recorded SLC hits are caused by single muons traversing the IceTop tanks.

The muon response model accounts for the trajectory of the muons through the IceTop tanks and the finite detector resolution. Although the chance for two simultaneous muon hits is very small at large radii, charge distributions for up to three simultaneous muon hits are considered. The default EM model (\emph{EM1}) assumes that the signals approximately follow a power-law. However, in order to allow small deviations from a simple power-law as a function of lateral distance, a second EM model (\emph{EM2}) is assumed and the differences between the models are included as systematic uncertainty in the final results. The contribution from accidental coincident background is modelled according to a Poisson distribution based on data taken in an off-time window before the EAS front arrives. The likelihood fit finds the mean number of muons per event $\langle N_\mu \rangle$ which is then divided by the cross-sectional area of the tanks projected onto a plane perpendicular to the shower axis to yield the muon density at a given location, $\rho_\mu(r)$.

\begin{wrapfigure}{r}{0.55\textwidth}
\vspace{-.5em}

\mbox{\hspace{-1em}\includegraphics[width=0.6\textwidth]{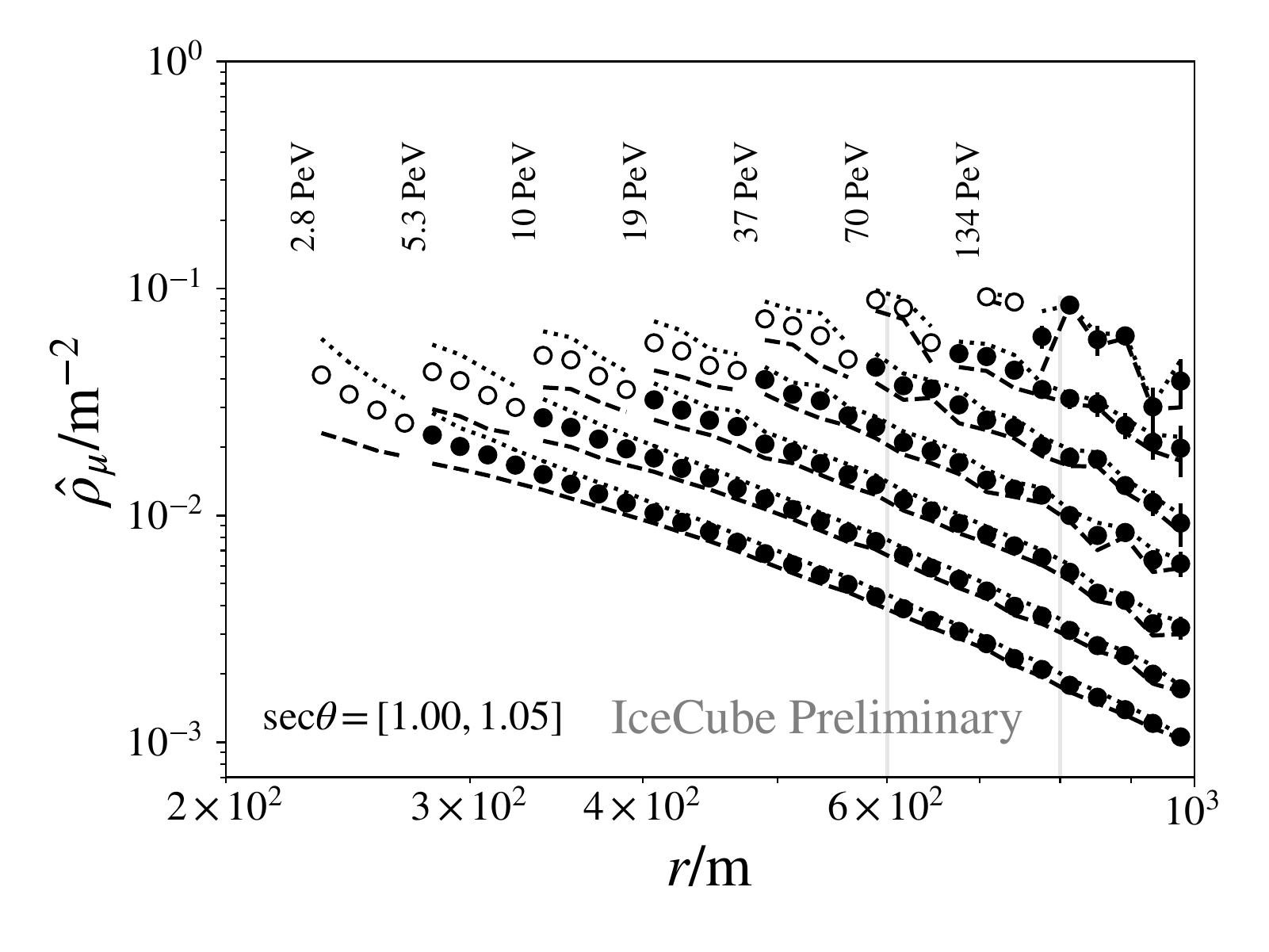}}
  \vspace{-2.25em}
  
\caption{The raw reconstructed muon densities, $\hat\rho_\mu$, as a function of lateral distance for seven energy bins. The lines indicate the systematic uncertainty associated to the function used to model signals with no muons. Filled markers correspond to lateral distances where more than 80\% of signals are SLC hits only.}
\label{fig:data_rho_ldf}
\vspace{-2.em}

\end{wrapfigure}

\begin{figure}
  \centering
  \vspace{-2em}
  
  \mbox{\hspace{-1em}\subfloat[]{%
  \label{fig:correction_sibyll}%
    \includegraphics[width=0.5\textwidth,trim=0.0 1.em 0.0 0.0, clip]{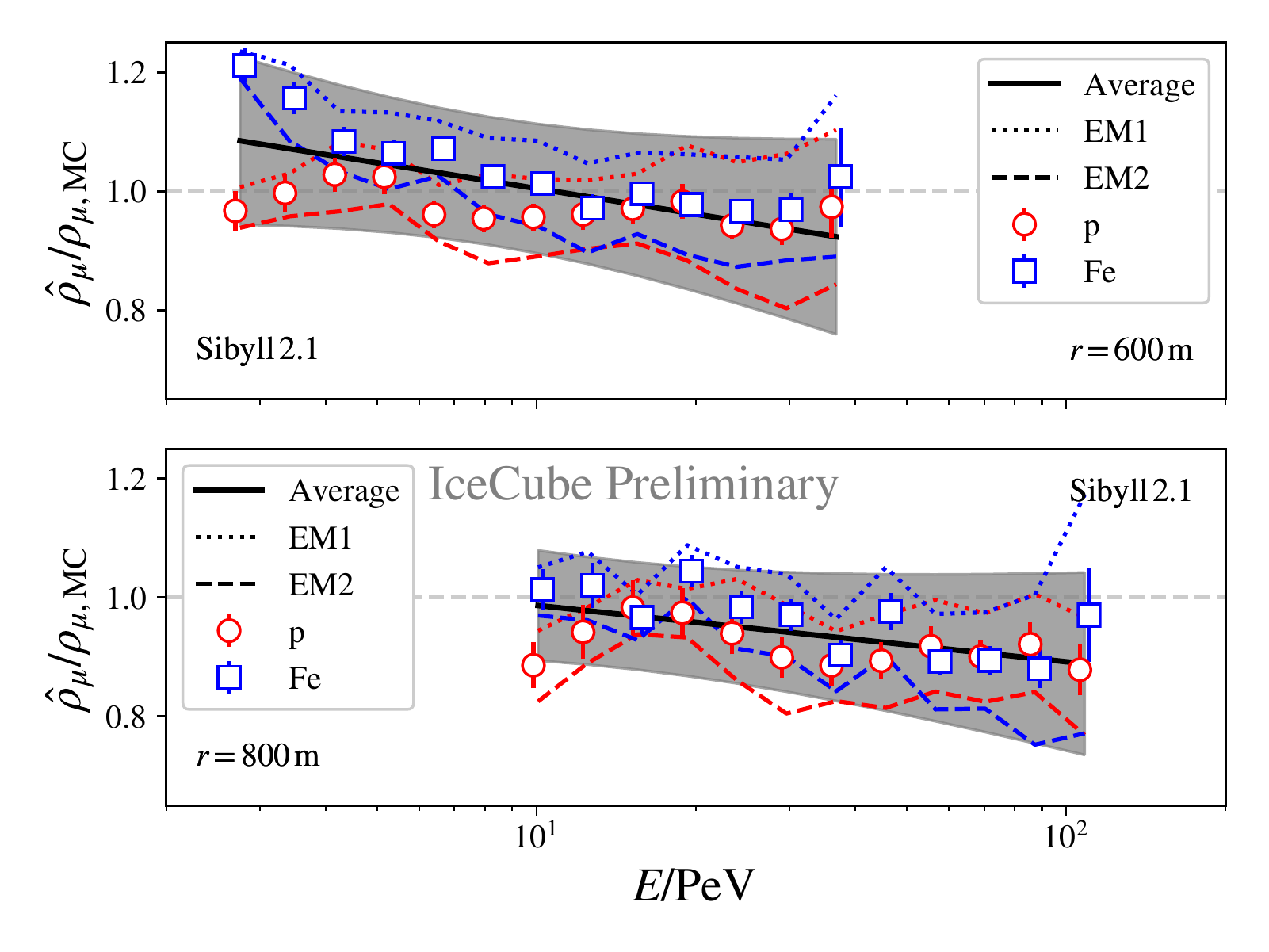}%
  }
  \subfloat[]{%
  \label{fig:correction_epos}%
    \includegraphics[width=0.5\textwidth,trim=0.0 1.em 0.0 0.0, clip]{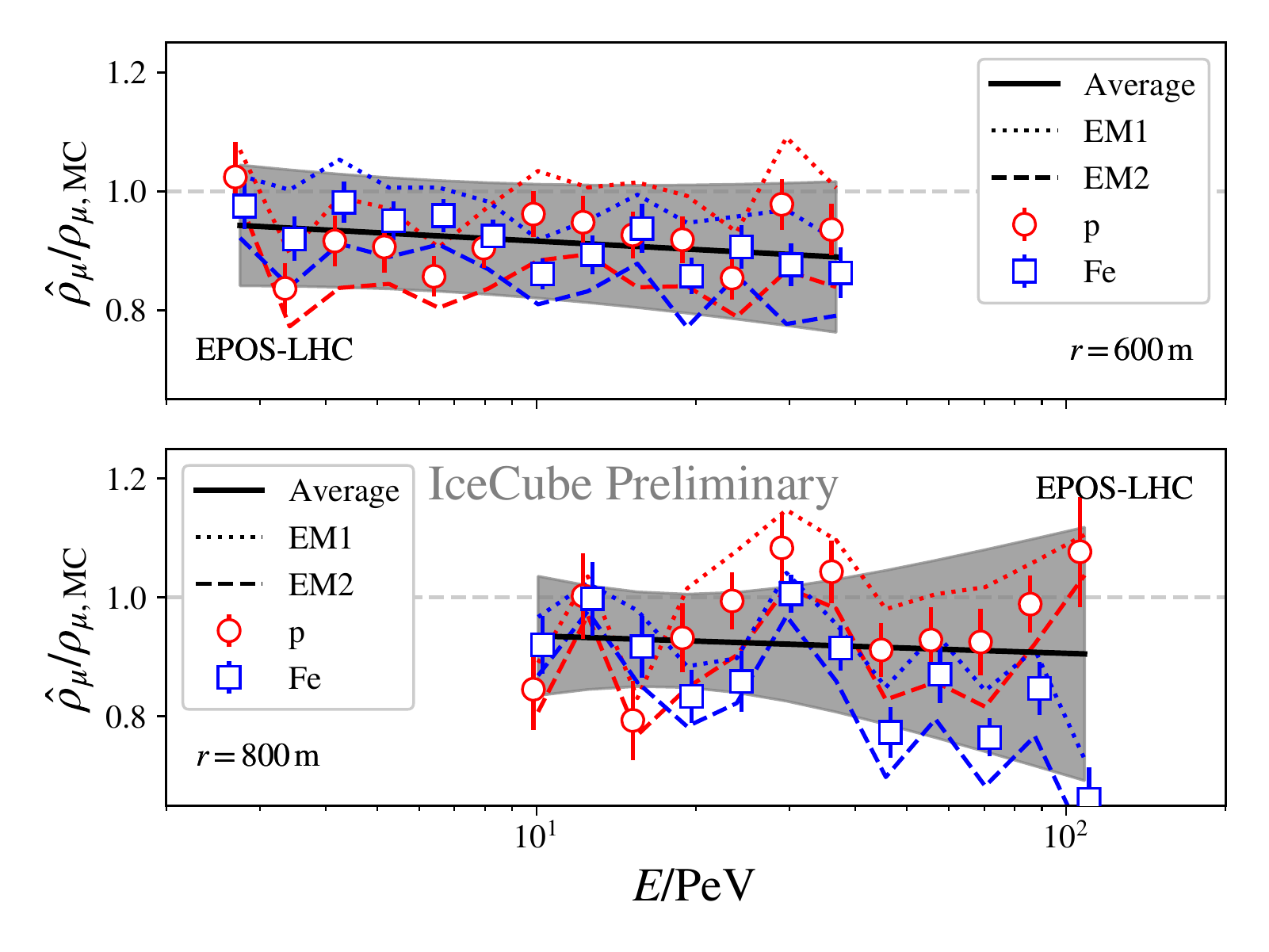}%
  }}\\
  
  \vspace{-1em}
  \mbox{\hspace{-1em}\subfloat[]{%
  \label{fig:correction_qgsjet}%
    \includegraphics[width=0.5\textwidth,trim=0.0 1.em 0.0 0.0, clip]{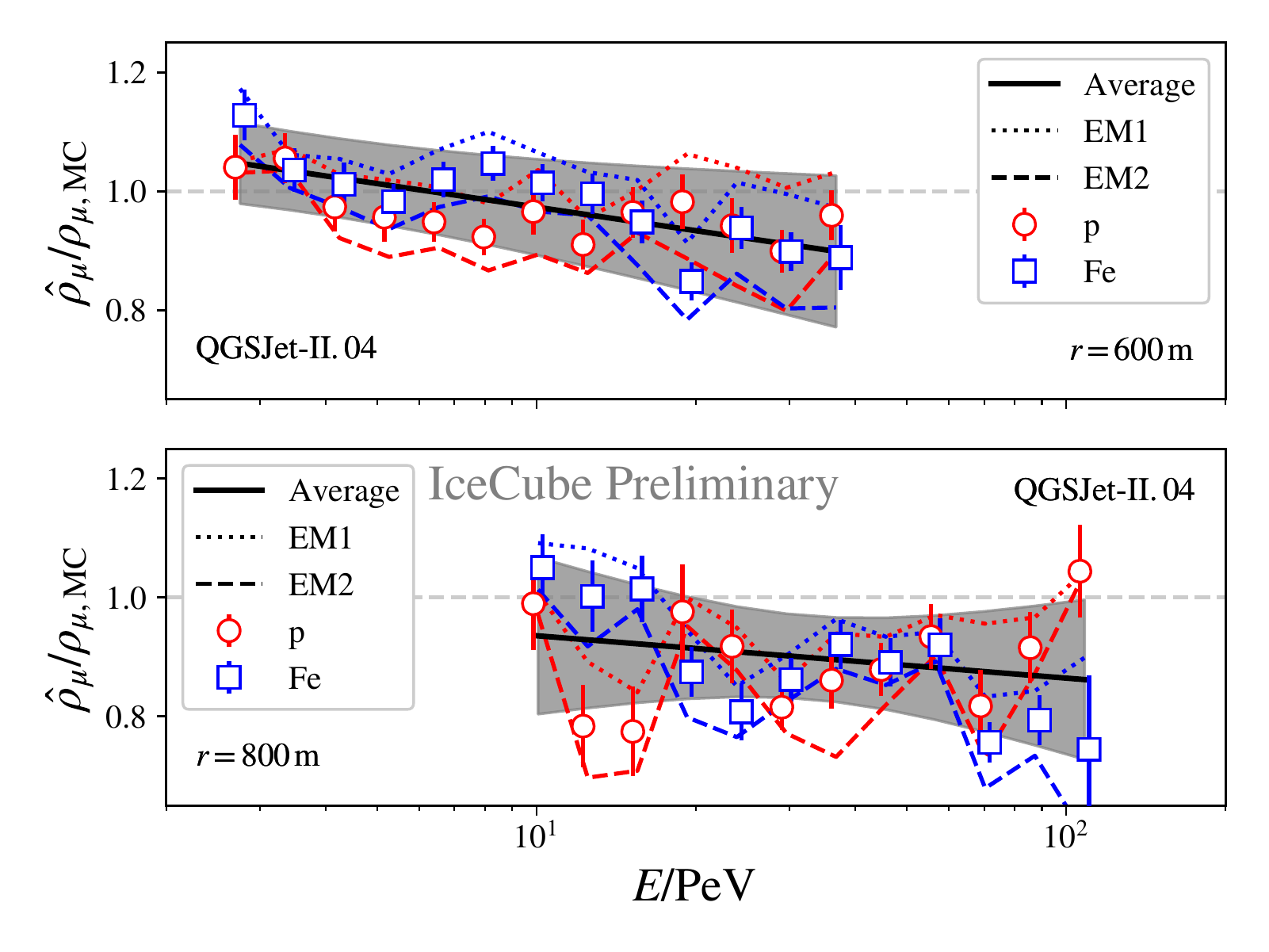}%
  }
  \subfloat[]{%
  \label{fig:correction_avg}%
    \includegraphics[width=0.5\textwidth,trim=0.0 1.em 0.0 0.0, clip]{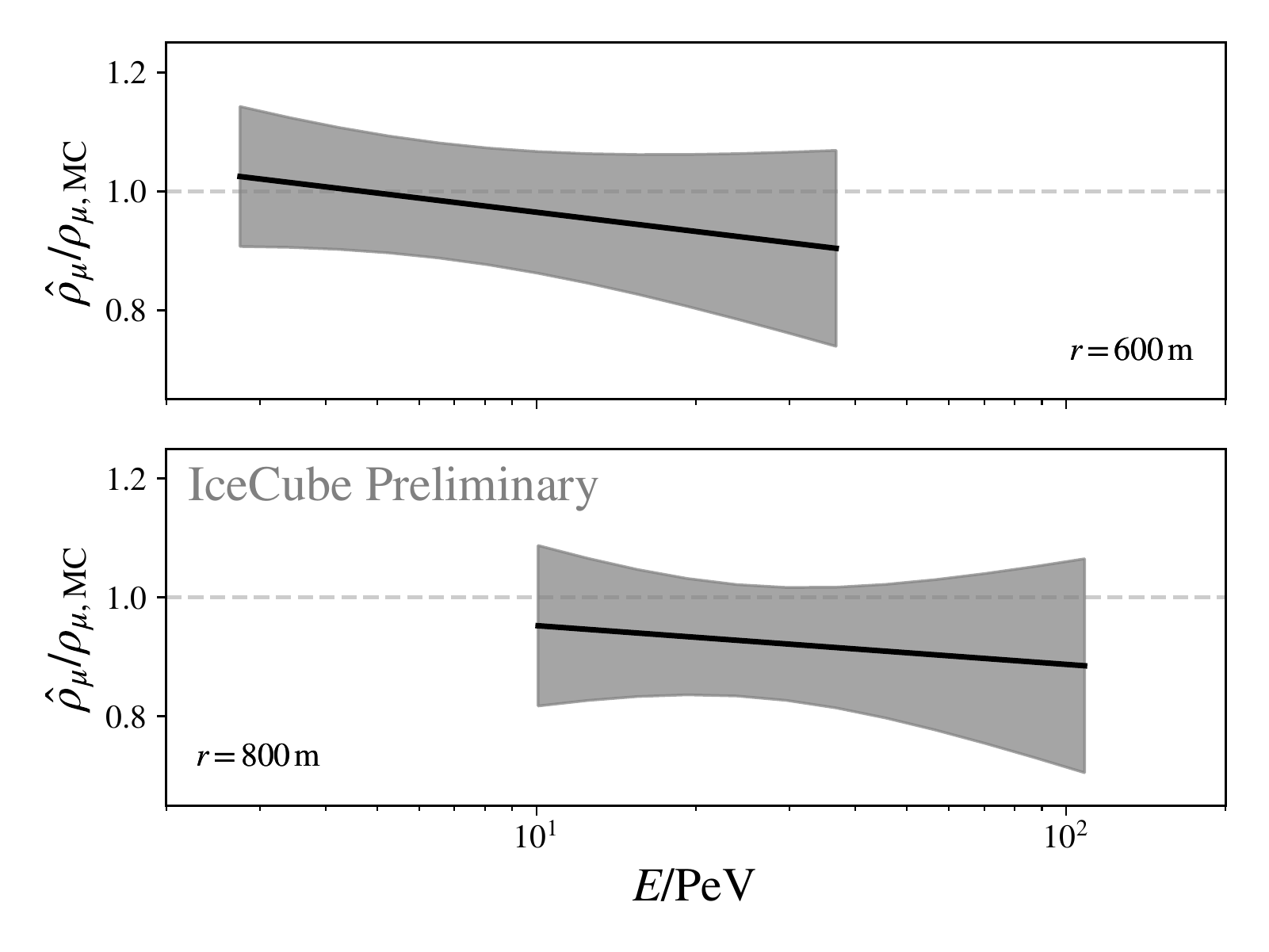}%
  }}
  \caption{Ratio of raw reconstructed over true muon density at \SI{600}{m} and \SI{800}{m} lateral distance, as a function of reconstructed energy, for simulated air showers using the hadronic interaction models Sibyll~2.1 (a), EPOS-LHC (b), and QGSJet-II.04 (c). Squares and circles correspond to iron and proton simulated showers, dotted and dashed lines indicate the ratios obtained using two different EM models. Corresponding fits are shown as solid lines where the grey band represents the corresponding uncertainties. Figure (d) shows the corresponding average ratios for these three hadronic models.}
  \label{fig:correction}%
  \vspace{-2em}
  
\end{figure}

The resulting \emph{raw} reconstructed muon densities, $\hat\rho_\mu(r)$, are shown in \cref{fig:data_rho_ldf}. These distributions are fit to interpolate the raw muon densities at radial distances of \SI{600}{m} and \SI{800}{m}. However, a small deviation of the raw reconstructed muon density from the truth is observed in simulations. This deviation is corrected by multiplying a correction factor to the raw density, $\hat\rho_\mu(r)$. %, as described in the next section. 
The correction is determined by dividing the reconstructed muon density in simulations by the true muon density. The resulting ratios obtained from proton and iron shower simulations using CORSIKA~\cite{CORSIKA_Heck} based on the hadronic interaction models Sibyll~2.1~\cite{Ahn:2009wx}, EPOS-LHC~\cite{Pierog:2013ria}, and QGSJet-II.04~\cite{Ostapchenko:2010vb}, are shown in \cref{fig:correction}. Also shown are the average ratios for these three hadronic interaction models.

\vspace{-2em}

%\subsection{Monte-Carlo Correction}
%\label{sec:correction}
%When this analysis is applied to simulations, small deviations on the order of up to $10\%$ between the raw reconstructed and true muon densities are observed which must be corrected in order to derive the muon densities in EAS. This correction is determined by dividing the reconstructed muon density in simulations by the true muon density. The resulting ratios obtained from proton and iron shower simulations using CORSIKA~\cite{CORSIKA_Heck} based on the hadronic interaction models Sibyll~2.1~\cite{Ahn:2009wx}, EPOS-LHC~\cite{Pierog:2013ria}, and QGSJet-II.04~\cite{Ostapchenko:2010vb}, are shown in \cref{fig:correction}.

\newpage

The inverse of this ratio is used as an \emph{MC correction factor} to adjust the result in reconstructed data. However, as shown in \cref{fig:correction}, the correction factor depends on the mass composition of the sample, since iron primaries require a larger correction. The actual composition is unknown, so the correction factor applied to the data is the average of the proton and iron factors, with a systematic uncertainty of half of the difference. A linear fit to this average yields the correction factors for each hadronic model, depicted as a black lines in \cref{fig:correction}. There are three contributions to the uncertainty in the correction factor, depicted as a grey band: the electromagnetic signal model used (EM1/EM2), the assumed mass composition, and the statistical uncertainty in the fit. All these uncertainties appear as systematic uncertainties in the resulting muon density and they are further discussed in the following. 

%\subsection{Systematic uncertainties}

The systematic uncertainty in this analysis arises from four main sources which are described in detail in Ref.~\cite{Aartsen:2021Muons}. The uncertainty in the energy determination causes a systematic uncertainty of about $7\%$ in the muon density because of the correlation between energy and muon number. In addition, the effect of the EM model assumption in the likelihood fit of the signal model introduces an uncertainty of up to about $10\%$ which is estimated from the differences obtained using the two electromagnetic models EM1 and EM2, as shown in \cref{fig:correction}. As also shown in \cref{fig:correction}, the correction depends on the composition assumption in MC and the corresponding uncertainty is estimated to be half the difference of the correction for proton and iron. The limited statistics of the simulations also introduces an uncertainty which is included in the uncertainty when applying the MC correction. For the results derived using the average correction in \cref{fig:correction_avg}, the model differences are also accounted for in the uncertainties.%In the following, systematic uncertainties are depicted by brackets around the central data points.

\section{Results}

\begin{wrapfigure}{r}{0.55\textwidth}
\vspace{-1.8em}

\mbox{\hspace{-.5em}\includegraphics[width=0.59\textwidth]{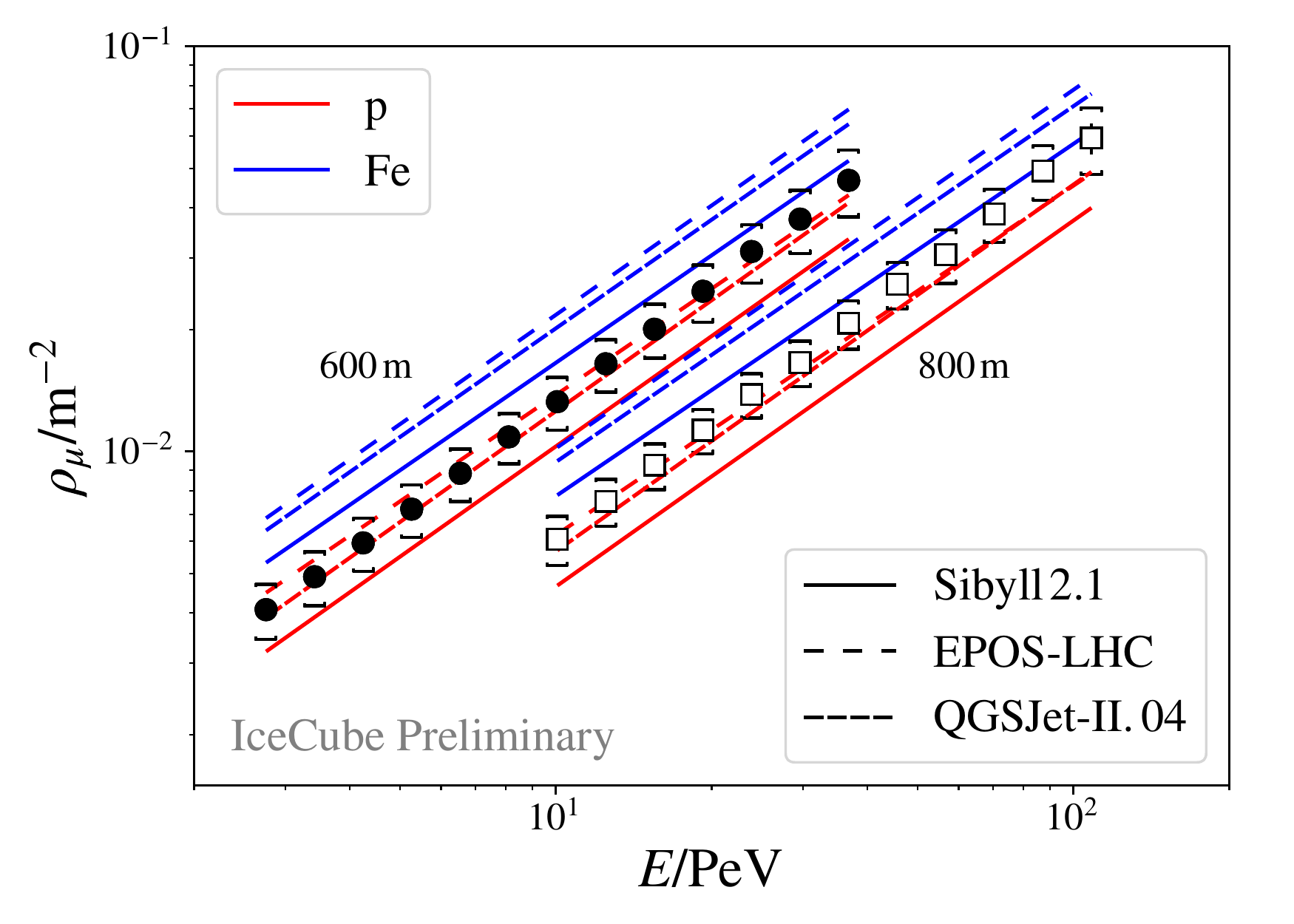}}
\vspace{-2em}
  
\caption{Measured muon density at \SI{600}{m} (solid circles) and \SI{600}{m} (white squares) lateral distance after applying the average correction from \cref{fig:correction_avg}. Error bars indicate the statistical, brackets the systematic uncertainty. Shown for comparison are the corresponding simulated densities for proton and iron (red and blue lines).}
\label{fig:rho_mu_final_avg}
\vspace{-1em}

\end{wrapfigure}

\begin{figure}
  \centering
  \vspace{-1.em}
  
  \mbox{\hspace{-1em}
    \includegraphics[width=0.482\textwidth,trim=0.0 -1.em 0.0 0.0, clip]{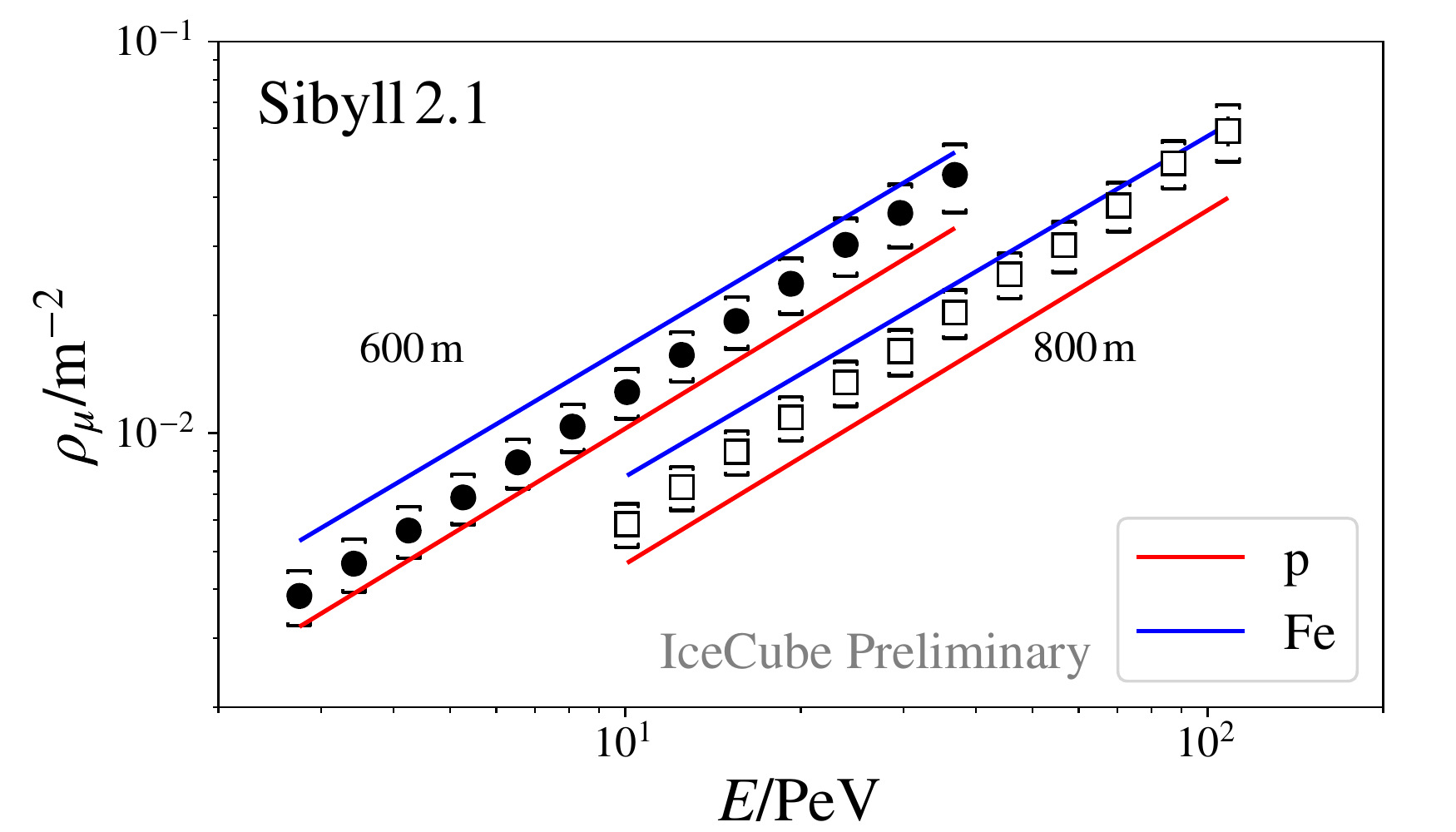}
    \includegraphics[width=0.535\textwidth,trim=0.0 2.em 0.0 0.0, clip]{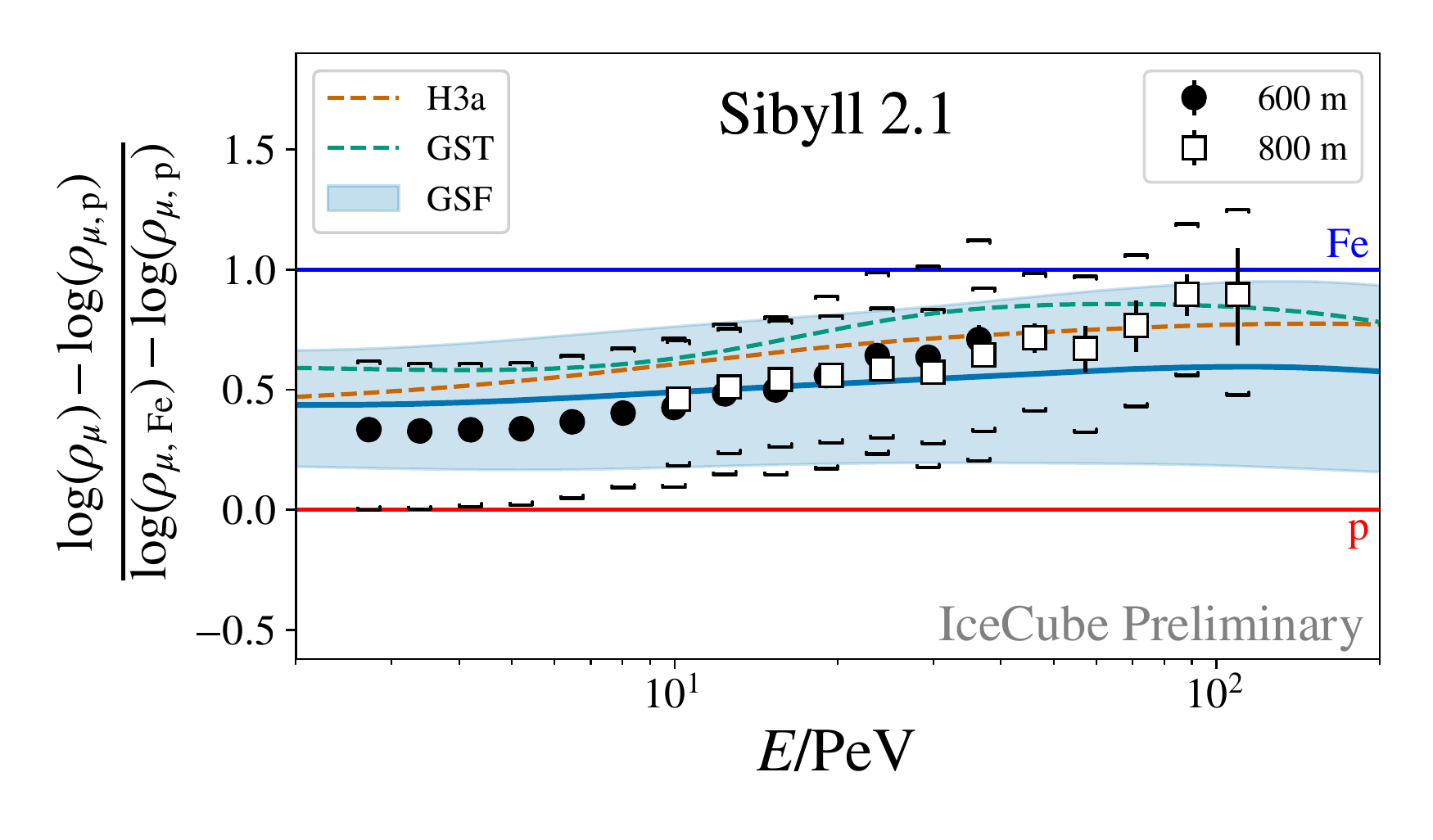}
    }
  \vspace{-1.6em}
  
  \mbox{\hspace{-1em}
    \includegraphics[width=0.482\textwidth,trim=0.0 -1.em 0.0 0.0, clip]{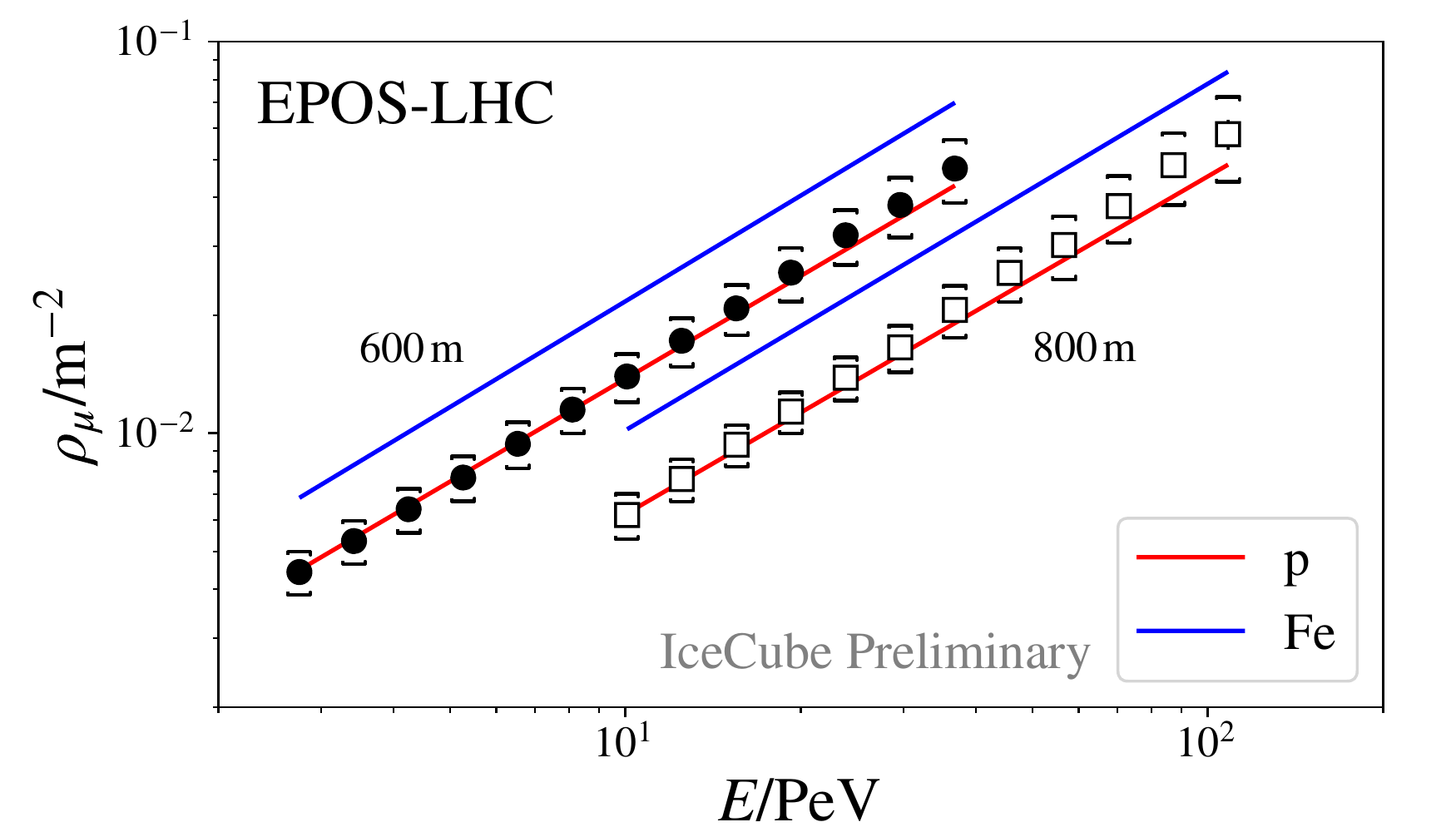}
    \includegraphics[width=0.535\textwidth,trim=0.0 2.em 0.0 0.0, clip]{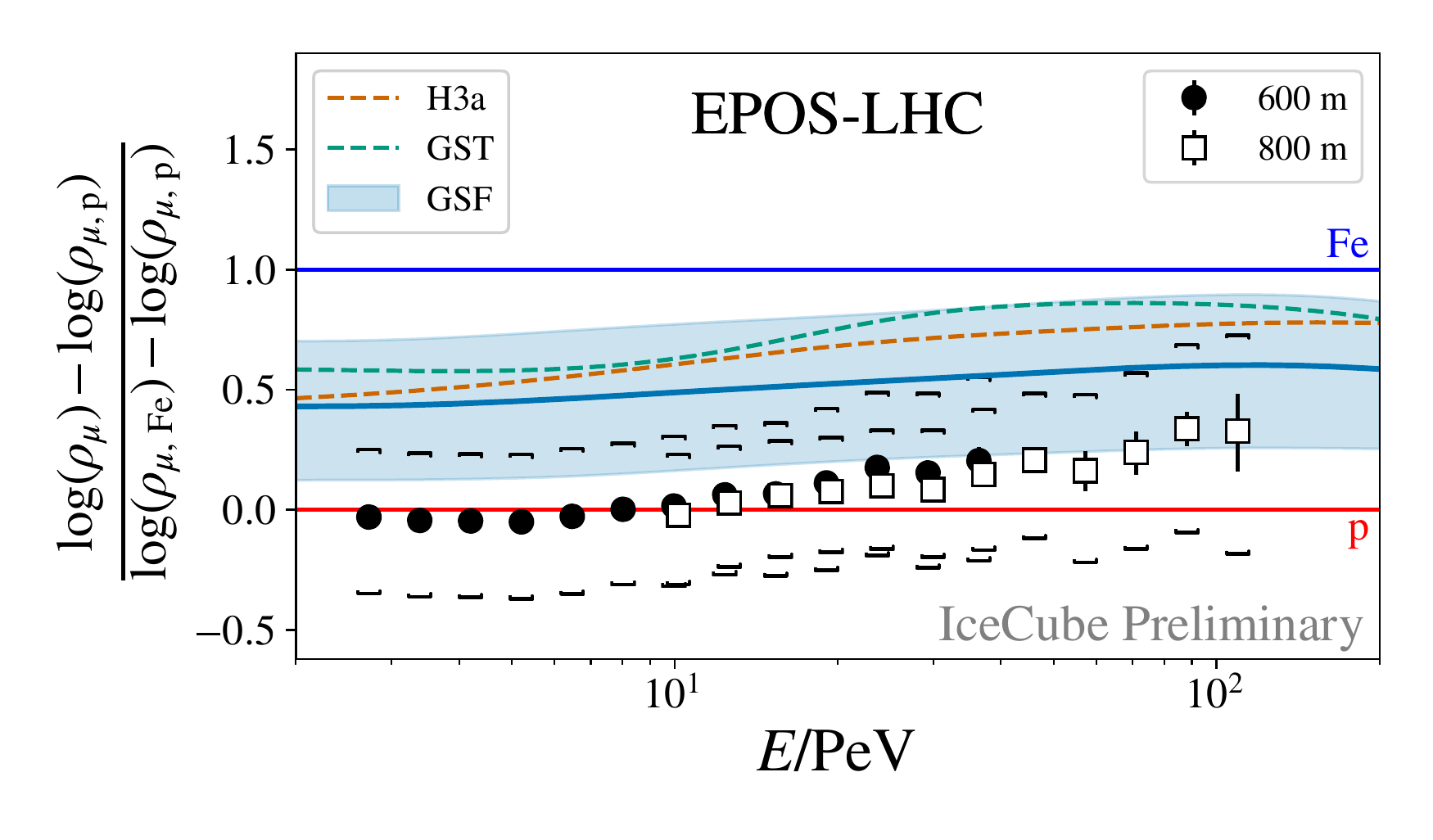}
  }
  \vspace{-1.6em}
  
  \mbox{\hspace{-1em}
    \includegraphics[width=0.482\textwidth,trim=0.0 -1.em 0.0 0.0, clip]{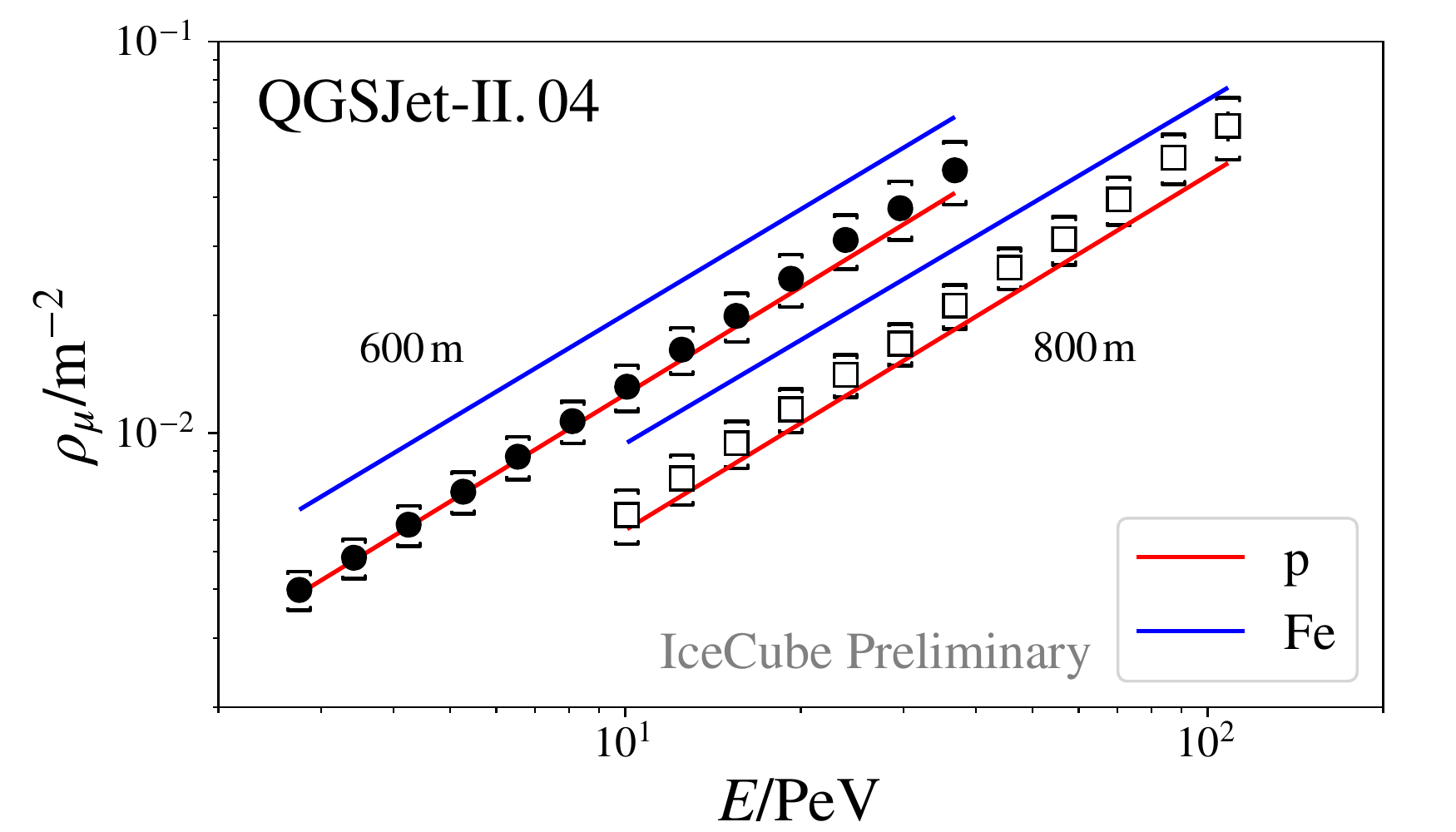}
    \includegraphics[width=0.535\textwidth,trim=0.0 2.em 0.0 0.0, clip]{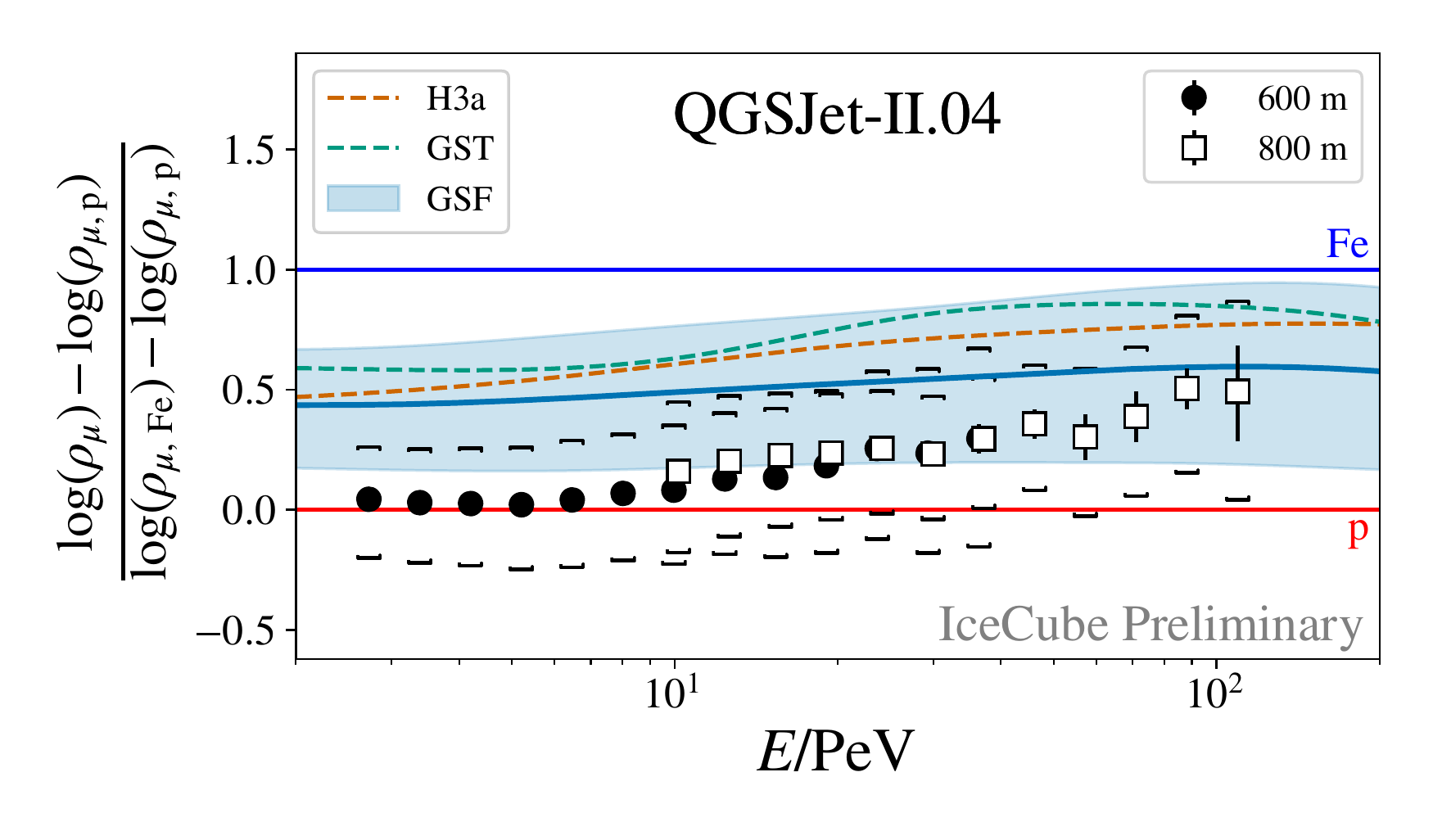}
  }
  \vspace{-1.5em}
  
  \caption{The final muon density measured in IceTop compared to predictions from the hadronic interaction models Sibyll~2.1 (top), EPOS-LHC (center), and QGSJet-II.04 (bottom). The figures on the right show the corresponding z-values, as defined in \cref{eq:z-values}, as well as expectations from the cosmic ray flux models GSF (with error band), GST, and H3a. The error bars represent the statistical, brackets systematic uncertainties.}
  \label{fig:rho_mu_final}%
  %\vspace{-1.em}
  
\end{figure}

The resulting muon densities at lateral distances of \SI{600}{m} and \SI{800}{m} for EAS energies from \SI{2.5}{PeV} to \SI{40}{PeV} and \SI{9}{PeV} to \SI{120}{PeV}, respectively, after applying the average correction factors from \cref{fig:correction_avg}, are shown in \cref{fig:rho_mu_final_avg}. The predictions from simulations based on the hadronic interaction models Sibyll~2.1, EPOS-LHC, and QGSJet-II.04 for proton and iron shower are shown as red and blue lines. The corresponding muon densities after applying the individual correction factors from \cref{fig:correction_sibyll,fig:correction_epos,fig:correction_qgsjet} for each hadronic interaction model separately are shown in \cref{fig:rho_mu_final} (left). Within their uncertainties, the measured muon densities are between the model predictions assuming proton and iron primaries. However, for the post-LHC models the predictions suggest a very light composition at the lowest energies below \SI{10}{PeV}, in particular using EPOS-LHC, which appears to be in tension with other experimental results~\cite{Kampert:2012mx,IceCube:2019hmk,Apel:2013dga}.

In order to compare the measured muon densities to predictions from different hadronic interaction models with certain cosmic ray flux assumptions in more detail, we define the quantity
\begin{equation}
\label{eq:z-values}
z=\frac{\log(\rho_{\mu})-\log(\rho_{\mu,\mathrm{p}})}{\log(\rho_{\mu,\mathrm{Fe}})-\log(\rho_{\mu,\mathrm{p}})}\; ,
\end{equation}
where $\rho_{\mu}$ is the measured muon density while $\rho_{\mu,\mathrm{p}}$ and $\rho_{\mu,\mathrm{Fe}}$ are the muon densities obtained from simulated proton and iron showers, respectively. The resulting distributions are shown in \cref{fig:rho_mu_final} (right), compared to predictions assuming the GSF~\cite{Dembinski:2015xtn}, GST~\cite{Gaisser:2013bla}, and H3a~\cite{Gaisser:2011cc} cosmic ray flux models. While Sibyll~2.1 describes the expected behaviour fairly well, at least up to energies of about \SI{50}{PeV}, the post-LHC models predict more muons, yielding a very light composition. This is in tension with the model predictions which are based on and are within their uncertainties in agreement with experimental observations of the cosmic ray mass composition.% (\emph{e.g.} Refs.~\cite{IceCube:2019hmk,Apel:2013dga}).

\section{Conclusions}

We have presented a measurement of the density of GeV muons at lateral distances of \SI{600}{m} and \SI{800}{m} for EAS energies from \SI{2.5}{PeV} to \SI{40}{PeV} and \SI{9}{PeV} to \SI{120}{PeV} in IceTop at an atmospheric depth of about $690\,\mathrm{g/cm}^2$. While the measured muon densities agree within their uncertainties with predictions based on Sibyll~2.1, the post-LHC models predict too large average muon densities assuming realistic cosmic ray flux models consistent with experimental data. While the model-tuning to LHC data improves the agreement with measurements at the highest EAS energies our results suggest that the muon densities in post-LHC models are not correct for primary energies below approximately \SI{100}{PeV}. However, a systematic shift in the reconstructed primary energy can also cause an apparent discrepancy in muon density. %A $20\%$ shift in the energy scale, for example, translates to an apparent shift of about 0.5 in the z-value, shown in \cref{fig:rho_mu_final}. 
For this reason, a detailed comparison of results from multiple observatories across all energies is needed to understand the production of GeV muons in air showers, as shown in Refs.~\cite{Dembinski:2019uta,Soldin:2021WHISP}. 

%Connection to the \emph{Muon Puzzle} and WHISP~\cite{Dembinski:2019uta,Soldin:2021WHISP} and connection to LHC physics~\cite{Albrecht:2021yla,Dembinski:2021LHC}. 

Future coincident measurements with the in-ice detector of IceCube and IceTop enable the simultaneous detection of GeV and TeV muons in EAS which provides spectral information and will help to further test hadronic interaction models~\cite{DeRidder:20174n,Verpoest:2021ICRC}. In addition, new surface extensions of the IceCube Neutrino Observatory~\cite{Haungs:2019ylq} in the context of IceCube-Gen2~\cite{Aartsen:2020fgd} will significantly increase the phase space of muon measurements towards larger zenith angles and higher EAS energies. This will allow measurements of the angular distribution of the atmospheric muon flux, up to EeV air shower energies, closing the gap towards EAS experiments at the highest energies.

\bibliographystyle{ICRC}
\bibliography{references}

%\clearpage

\section*{Full Author List: IceCube Collaboration}

% \noindent \textbf{Note comment afterwards:} Collaborations have the possibility to provide an authors list in xml format which will be used while generating the DOI entries making the full authors list searchable in databases like Inspire HEP. For instructions please go to icrc2021.desy.de/proceedings or contact us under icrc2021proc@desy.de.\\

% \scriptsize
% \noindent
% first.author$^1$, 
% second.author$^2$, 
% third.author$^3$ % .... more names
% and 
% last.author$^{n}$ \\

% \noindent
% $^1$first.affiliation.
% $^2$second.affiliation. % .... more affiliation
% $^{m}$last.affiliation.

\scriptsize
\noindent
R. Abbasi$^{17}$,
M. Ackermann$^{59}$,
J. Adams$^{18}$,
J. A. Aguilar$^{12}$,
M. Ahlers$^{22}$,
M. Ahrens$^{50}$,
C. Alispach$^{28}$,
A. A. Alves Jr.$^{31}$,
N. M. Amin$^{42}$,
R. An$^{14}$
% ...
\\

\noindent
% ...
$^{12}$ Universit{\'e} Libre de Bruxelles, Science Faculty CP230, B-1050 Brussels, Belgium \\
% ...
$^{14}$ Department of Physics and Laboratory for Particle Physics and Cosmology, Harvard University, Cambridge, MA 02138, USA \\
% ...
$^{17}$ Department of Physics, Loyola University Chicago, Chicago, IL 60660, USA \\
$^{18}$ Dept. of Physics and Astronomy, University of Canterbury, Private Bag 4800, Christchurch, New Zealand \\
% ...
$^{22}$ Niels Bohr Institute, University of Copenhagen, DK-2100 Copenhagen, Denmark \\
% ...
$^{28}$ D{\'e}partement de physique nucl{\'e}aire et corpusculaire, Universit{\'e} de Gen{\`e}ve, CH-1211 Gen{\`e}ve, Switzerland \\
% ...
$^{31}$ Karlsruhe Institute of Technology, Institute for Astroparticle Physics, D-76021 Karlsruhe, Germany  \\
% ...
$^{42}$ Bartol Research Institute and Dept. of Physics and Astronomy, University of Delaware, Newark, DE 19716, USA \\
% ...
$^{50}$ Oskar Klein Centre and Dept. of Physics, Stockholm University, SE-10691 Stockholm, Sweden \\
% ...
$^{59}$ DESY, D-15738 Zeuthen, Germany \\
% ...

\subsection*{Acknowledgements}

\noindent
USA {\textendash} U.S. National Science Foundation-Office of Polar Programs,
U.S. National Science Foundation-Physics Division,
U.S. National Science Foundation-EPSCoR,
Wisconsin Alumni Research Foundation,
Center for High Throughput Computing (CHTC) at the University of Wisconsin{\textendash}Madison,
Open Science Grid (OSG),
Extreme Science and Engineering Discovery Environment (XSEDE),
Frontera computing project at the Texas Advanced Computing Center,
U.S. Department of Energy-National Energy Research Scientific Computing Center,
Particle astrophysics research computing center at the University of Maryland,
Institute for Cyber-Enabled Research at Michigan State University,
and Astroparticle physics computational facility at Marquette University;
Belgium {\textendash} Funds for Scientific Research (FRS-FNRS and FWO),
FWO Odysseus and Big Science programmes,
and Belgian Federal Science Policy Office (Belspo);
Germany {\textendash} Bundesministerium f{\"u}r Bildung und Forschung (BMBF),
Deutsche Forschungsgemeinschaft (DFG),
Helmholtz Alliance for Astroparticle Physics (HAP),
Initiative and Networking Fund of the Helmholtz Association,
Deutsches Elektronen Synchrotron (DESY),
and High Performance Computing cluster of the RWTH Aachen;
Sweden {\textendash} Swedish Research Council,
Swedish Polar Research Secretariat,
Swedish National Infrastructure for Computing (SNIC),
and Knut and Alice Wallenberg Foundation;
Australia {\textendash} Australian Research Council;
Canada {\textendash} Natural Sciences and Engineering Research Council of Canada,
Calcul Qu{\'e}bec, Compute Ontario, Canada Foundation for Innovation, WestGrid, and Compute Canada;
Denmark {\textendash} Villum Fonden and Carlsberg Foundation;
New Zealand {\textendash} Marsden Fund;
Japan {\textendash} Japan Society for Promotion of Science (JSPS)
and Institute for Global Prominent Research (IGPR) of Chiba University;
Korea {\textendash} National Research Foundation of Korea (NRF);
Switzerland {\textendash} Swiss National Science Foundation (SNSF);
United Kingdom {\textendash} Department of Physics, University of Oxford.

\end{document}